# Context-Aware Middleware: A Review

**Hamed Vahdat-Nejad**

Pervasive Computing lab, Computer Engineering Department, University of Birjand

vahdatnejad@birjand.ac.ir

**Abstract** During previous years several studies have introduced the concept of "context-aware middleware" and also proposed solutions under this title; however, these systems are different in functionality. In this chapter, context-aware middleware is investigated from the standpoint of functional and non-functional requirements. Afterward, some well-known middleware systems are reviewed and, finally, open research directions as well as conclusion remarks are presented.

## 1. Introduction

Context-aware applications have been considered as the building blocks of the pervasive computing paradigm. The main challenge involved with implementing a context-aware application is discovering and obtaining reliable contextual information from the environment. Designing and implementing a stand-alone context-aware application is a lengthy and sophisticated process, suffering from the following issues:

- The design and development procedure takes an excessive amount of time and cost.
- A majority of context-aware applications reside on devices with limited memory, computation, and communication capabilities such as mobile phone, PDA, tablet, and wearable computer. Hence, it is not reasonable to execute a massive program on them.

As a result, a research direction which spans design and deployment of a supporting system for developing context-aware applications has arisen. These systems, which are generally referred to as middleware, have been investigated under different titles including context acquisition and dissemination [1], distribution [2], management [3], toolkit [4] and even engine [5]. There are several reasons for necessity of middleware introduction in pervasive systems:

- The Infrastructure of pervasive computing consists of many non-dedicated devices as well as sensors with limitations in memory, storage, computation and availability. Managing the vast number of devices and sensors and storing and processing the enormous volume of generated contextual data require extensive system support.
- In an environment with a variety of context-aware applications, many of single context elements are needed by different applications; each programmer may develop context provider software components for obtaining their required sophisticated context types. Therefore, designing stand-alone context-aware applications is inefficient in terms of reusability criterion.
- From the initial categorization of context types a decade ago, new dimensions of context have arisen such as social and urban context. Furthermore, context reasoning, mining and other techniques have led to the emergence of new sophisticated context types. These remarks motivate the design of an open cooperative system in which different entities could introduce new context elements and share their derived and estimated values.
- Finding and composing available context-aware services is difficult for application developers and needs system support [6].

In general, middleware is a software layer, which by residing between the operating system and application layer in each node, provides new capabilities and facilitates development of applications. Utilizing middleware leads to design of well-architected distributed systems [7][8]. However, a middleware platform for pervasive computing or in particular, context-aware middleware, is different in certain aspects from traditional middleware in distributed systems.

To continue, section 2 provides requirements of context-aware middleware. Section 3 is dedicated to surveying some well-known proposals for context-aware middleware and finally, section 4 discusses open research directions and concludes the chapter.

## 2. Context-aware middleware

In general, requirements of a middleware platform are of two types [9]: (1) functional, which involves tasks that should be performed by the middleware, and (2) non-functional, which consists of the quality attributes (such as performance, availability, usability, extensibility, etc) that should be satisfied by the system. In the following subsections, functional and non-functional requirements of general context-aware middleware are discussed.

## *2-1 Functional requirements*

Context-aware middleware should generally address the following functional requirements:

- *Context acquisition*: Middleware should be able to discover context sources available in the environment. Usually, context-aware middleware systems provide registration mechanisms for context sources to declare their contextual capacities. Because most sensors and devices have a limited storage memory, it is the duty of middleware to gather and store contextual data.

- *Aggregation*: In wireless sensor networks, a diverse range of sensors produce a large amount of raw data. However, they aim to provide a few high-level pieces of contextual information. For example, numerous sensors may be set up inside a jungle to collaboratively detect fires and floods. Storing all these raw contextual data is not reasonable. Instead, context aggregation is utilized to obtain low-volume meaningful information for storage and transmission.

- *Quality of context (QoC) assessment*: Context is inaccurate and uncertain because of three aspects: inaccuracy of sensors, inaccuracy of reasoning algorithms, and dynamic and temporal nature of context. On the other hand, diverse applications require different levels of quality for their contextual needs. Inaccurate or unreliable values of context may result in serious problems for example, for applications available in the pervasive healthcare or elderly-care domain. In general, many applications need a specific minimum level of quality for their contextual requirements, with respect to which the context, produced by heterogeneous sources all over the environment, should be evaluated. Quality of context deals with assessing and measuring quality of a context element against application requirements. Besides, sometimes there is more than one source for a single context element and their provided values are different. In this case, middleware is responsible for resolving this conflict. Finally, the recently introduced "context provenance" [10] notion, which consists of mechanisms for tracking the origin of context, lies in this part. Provenance can be used for assessing quality and reliability of context.

- *Modeling*: Raw contextual data produced by sources (e.g. sensors) should be modeled and transformed to meaningful information to be usable by applications. Because context (e.g. a moving person) involves multi-dimensional time series data, traditional approaches like key/value are not effective for modeling. Many prior studies have been performed on context modeling [11]. Main approaches to context modeling are object-role based, spatial models and ontology-based approaches [11].

- *Reasoning*: Sensors can only measure simple context types. It is not possible to directly measure a high-level context such as activity or fall of a person through a sensor. In these situations, a reasoning component is exploited to derive the high-level context type from low-level ones.

- *Context dissemination*: Interaction between context-aware middleware and applications is performed via context dissemination mechanism. In fact, a context-aware application uses context dissemination mechanism provided by the middleware to obtain its required context. Context dissemination mainly involves event-driven and query-based approaches. In the event-driven approach, when an event (context update) occurs, middleware publishes the new value to the interested applications. In the query-based approach, middleware provides a query-based language for disseminating contextual information.

- *Service management*: In a smart environment diverse types of basic services exist. A service can be a software/hardware service for controlling a device (such as starting fire alarm, showing daily news on the screen), or a basic context-aware service (such as turning off lights in a building after everyone has left for the day), or even a complex context-aware application. Application developers usually search for these miniature service components and try to build their context-aware applications by composing suitable available services. Context-aware middleware should provide the functionality for service discovery and composition. Directory-based and DHT-based approaches are popular for service discovery.

- *Privacy protection*: Many of context types characterizing users are considered as private information by themselves and should not be openly disclosed [12]. However, a major part of context-aware applications rely on user's private information such as location, activity, health status, etc. The system should follow the policies of users when distributing their context among context-aware applications. Access control is widely used [4] [13] [14] for protecting user's contextual information from unauthorized parties. Pseudo-nymity and anonymity [15] are other solutions for privacy protection. In the first, users change their pseudonyms regularly to hide their identity. In the latter, for general location-based services users anonymously request for service.

## *2-2 Non-functional requirements*

Each non-functional requirement is associated with a software quality attribute. In the IEEE scientific expressions definition [16], quality is the "degree of which the system satisfies requirements of its users". Quality attributes are the factors and parameters that influence overall quality of software. Each system is constructed to satisfy specific quality attributes, where many of these attributes such as usability are generally important for all systems. In this subsection, we focus on specific quality attributes that should be considered in the design stage of a context-aware middleware system:

- *Expandability*: A typical pervasive computing environment consists of several domains. Typically, the number of domains increases, by joining other homes, offices, organizations, hospitals, urban, social, and user personal domains, during the runtime of the system. Furthermore, over time, new entities and context types are introduced. Therefore, middleware should be expandable from the viewpoint of domain, entity, and context.

- *Transparency*: Context-aware middleware should provide transparency for user's usage. Users typically interact with the middleware by utilizing context dissemination mechanisms. As a result, context dissemination should be transparent from two aspects: access and location. Context is provided and preprocessed by diverse programs, which are implemented using different languages, and residing on various platforms (e.g. mobile phone and Android OS, PC and Windows OS). Access transparency means that the programmer should utilize a common service, API or method for retrieving any of the context types. Location transparency indicates that the middleware should provide requested contextual information from anywhere (domain, server, etc) inside the environment without bothering the user with the location of the context provider, store, server, etc.

- *Reusability*: Users normally develop software modules to acquire implicit and high-level contextual information. However, these modules may be needed by a diverse range of programmers. Another kind of reusability is concerned with sharing context-aware service components implemented by application programmers. Context-aware middleware should provide an open framework for facilitating component reuse and share between programmers. Providing a uniform understandability of the components and a systematic strategy for component retrieval are the most important challenges that the middleware should overcome to realize component reusability [18].

- *Reliability:* Pervasive computing aims to help users in their daily tasks by offering everywhere every time services. Incorrect, inaccurate, early, and overdue services devastate the trust of the users in the system. According to the domain and aim of the middleware, different degrees of reliability are required. For some application domains, such as healthcare and collision avoidance in context-aware transportation systems, reliability is regarded as a critical factor. For example, an overdue detection of a person fall or an accident may result in loss of life.

## 3. A survey on context-aware middleware systems

In this section, we review some well-known context-aware middleware systems. For this, at first a framework for systematically studying the projects is provided. Then in the subsequent subsections, results of reviewing each project are discussed according to the framework.

### *3-1 Reviewing framework*

We review middleware systems from the standpoint of three aspects:
- *Overview*: From this viewpoint, we review the scope of the assumed environment of each project and structure of their middleware system.
- *Functional tasks*: Each middleware system supports some of the functional requirements that have been stated in the previous section.
- *Non-functional attributes*: Each middleware system satisfies some of the non-functional requirements that have been stated in the previous section.

From many projects related to context-aware middleware, we survey some well-known research studies including Context Toolkit [4], Gaia [13], Cobra [19], SOCAM [20] [21] [22], Awareness [23] [24] [25] [26] [27], SM4ALL [28] [29] [30], Feel@home [31] [14], Open [32], and CAMEO [33].

### *3-2 Overview*

***Scope of the environment***- Context Toolkit, Cobra, SOCAM, SM4ALL, and CAMEO investigate a single-domain pervasive environment. Context Toolkit assumes a general domain, but Cobra, SOCAM and SM4ALL consider a specific domain. Cobra is proposed for a meeting room. SOCAM and SM4ALL propose middleware for smart home domain. CAMEO investigates the mobile domain in which every user holds a mobile phone.

Among multiple-domain projects, Gaia assumes an active space, which consists of homes, offices, and meeting rooms. Awareness assumes four domains: mobile, home, office, and ad-hoc, and Feel@home initially considers three domains: home, office, and outdoor. Table I summarizes scope of these middleware systems.

*Architecture structure*- Structure of middleware architecture could be centralized, flat distributed, or hierarchal [17]. Context toolkit is based on a central main component known as discoverer; therefore, the architecture of the middleware is centralized. Gaia is based on central context service component, which is responsible for the main context management tasks. Cobra is based on multi-agent systems, in which a central context broker is responsible for context management tasks and is regarded as the main component of the system. SOCAM's main component is service locating service, which is designed centrally. In the Awareness project, the architecture of the middleware of the mobile, home and office domains [25] is centralized and the architecture of the middleware of the ad-hoc environment [24] is peer to peer; however, bridges make a flat distributed architecture for the overall middleware. SM4ALL makes use of several central registries involving processor registry, context type registry and publisher registry. Feel@home utilizes a hierarchical architecture in which GAS is the vertex, and domain context managers are the branches. CAMEO is designed as a single software package, which is distributed on user's mobile phones. The mobile phones use peer-to-peer paradigm for communication.

**Table I. Overview of the projects**

| Project | Scope of environment | Description | Structure |
|---|---|---|---|
| Context Toolkit | single-domain | general | Centralized |
| Gaia | multiple-domain | Active space involving homes, offices, and meeting rooms | Centralized |
| Cobra | single-domain | meeting room | Centralized |
| SOCAM | single-domain | smart home | Centralized |
| Awareness | multiple-domain | mobile, home, office, and ad-hoc | Distributed |
| SM4ALL | single-domain | smart home | Centralized |
| Feel@home | multiple-domain | home, office, and outdoor | Hierarchal |
| CAMEO | Single-domain | mobile domain | Distributed |

### *3-3 Functional tasks*

In this subsection, we look through the context-aware middleware systems from the viewpoint of their functional capabilities. At the end, summary of the investigation is presented within a table.

- *Context acquisition* - Context Toolkit is based on context widgets. Each widget is a software component responsible for gathering a specific context type from sensors. Widgets register themselves to the discoverer component in order to declare their contextual capacities. Widgets store sensed context and can provide a history of them to the interested entities. In Gaia, the context service uses a registry component, for context discovery. The registry maintains the information about all context sources in the environment. In Cobra, Context source discovery is accomplished via sensing the presence of Bluetooth MAC addresses. It then stores XML-based contextual information in a relational database. In SOCAM, Context providers gather raw contextual data from sources. There is one logical context database in each domain, which stores a history of contextual information of that domain. The service locating component plays the role of context discovery by providing a mechanism for context providers and interpreters to advertise their contextual capacities. The mechanism needs context providers to be registered into a service registry. Awareness utilizes registration service for context source discovery. For this, context producers are required to register themselves to the context broker components. Awareness partially supports context storage by introducing context storage engine. Such an engine could subscribe to some context producers and store published context. In SM4ALL, each sensor has a wrapper, which serves as a context provider by representing the device as a web service. The wrappers register themselves to the Publisher Registry (discovery). Storage is supported by introducing context persistence as an optional plug-in. In Feel@home, Context Wrappers gather raw data from sources and send it to the context aggregator, which in turn triggers JENA operations to store it. CAMEO introduces Device Context Provider, which is in charge of collecting context data derived from internal components of the mobile phone. It also supports storage by introducing history as an enrichment parameter for context elements.

- *Aggregation* - Among the surveyed projects, only SM4ALL supports aggregating low-level context types over a long period by introducing various types of context processing paradigms. Summarization and aggregation are among them. In the summarization paradigm, a specific context type is gathered over a period and summarized into a single value, e.g. the power usage of a fridge for each hour. Aggregation pattern gathers some low-level contextual information and infers and aggregates it to a single high-level context element. The hybrid of summarization and aggregation is utilized for aggregating raw contextual data generated by several sensors over a time interval. An example is gathering data of several sensors, which are monitoring health status of a person, and publishing a periodic high-level context- "User Healthy".

- *Quality of context assessment* - Among the projects, Awareness and SM4ALL provide functions for assessing QoC. In general, Awareness exploits the parameters of freshness, spatial resolution, temporal resolution and probability of correctness for assessing QOC. In SM4ALL, QOC has been considered as an optional attribute to all context types. It consists of three metrics: freshness, trust-worthiness, and precision. On the other hand, QOC evaluator is designed as an optional plug-in. CAMEO supports QoC by enriching the context model by several quality parameters including accuracy, freshness, cardinality, and dependencies between fact types.

  Cobra and SOCAM only perform a kind of conflict resolution without assessing QoC. In Cobra, a kind of simple conflict resolution is performed by the context broker via detecting and resolving inconsistent knowledge stored in the shared context model [19]. In SOCAM, the Context interpreter component, which involves a context Knowledge Base (KB), is responsible for performing conflict resolution by maintaining the consistency of context Knowledge Base and resolving conflicts.

- *Modeling* - Context Toolkit makes use of an object-oriented modeling scheme, which is performed by widgets. In Gaia, context modeling is based on first order logic and Boolean algebra. A 4-ary predicate structure is adopted from simple English clauses to represent context. Cobra makes use of RDF for modeling and representing context. In SOCAM Context providers perform modeling using Web Ontology Language (OWL). There is one logical context database in each domain, which stores context ontologies of that domain. In Awareness, local middleware of domains provides different mechanisms for context modeling. In SM4ALL, a central Context Type component is responsible for context modeling. It makes use of an object-oriented scheme, which declares a unique name and list of attributes for each context type. In Feel@home, Context Wrappers perform modeling by transforming the obtained raw data into context markups. CAMEO makes use of Context Modeling Language (CML) as an extension of the object-role-based model. It provides formal representation for denoting object types and fact types. A fact type denotes a relationship between two object types.

- *Reasoning* - Gaia supports limited context reasoning by using first order logic and Boolean algebra. Cobra utilizes OWL and rule-based inference for reasoning about high-level contextual information. In SOCAM the Context interpreter also involves reasoner component, which is responsible for deducing high-level context using logic reasoning. In Awareness context reasoning is performed by distributed context reasoners, which acquire low-level contextual information from various context producers and infer high-level contextual information. In SM4ALL aggregation pattern also plays the role of context reasoning by gathering some low-level contextual information and inferring it to a single high-level context element. Feel@home provides JENA component, which is based on the Jena Semantic Web package and OWL. It provides an inference engine that can infer high-level context. CAMEO makes use of CML reasoning technique, which is based on three-valued logic.

- *Context dissemination* - Context Toolkit provides a query-based mechanism for context dissemination. A context-aware application should query the discoverer to find the widget associated to the required context, and then should interact directly with the widget for subscribing to the context. In Gaia the context service provides a query-based dissemination approach for applications to query and register for their required context types. Cobra does not provide any context dissemination mechanism for independent context-aware applications. In SOCAM context dissemination is performed by service locating component via providing query mechanism for context-aware applications. Afterward, the context-aware applications find the context providers that present their contextual needs. Subsequently, they can directly obtain the required context via either query or event-driven (pull or push based) approaches. Awareness makes use of query and subscription approaches for context dissemination. In SM4ALL, context-aware applications acquire contextual needs by using context listeners. Each listener is associated with a context query and listens to the notifications of the middleware (dissemination). Feel@home utilizes publish-subscription paradigm for intra-domain context dissemination and query approach for global context dissemination. In CAMEO, each application should register to the middleware by specifying its required context element. During the registration, a unique identifier is assigned to the application. Afterwards, the middleware is responsible for notifying the application of context changes. In addition, CAMEO provides an application programming interface towards mobile social network applications. CAMEO also introduces the beaconing module, which implements periodical context exchange among one-hop neighbors.

- *Service management* - Among the surveyed projects, only Open (Extension of Feel@home) supports service management. In the Open framework, a programming toolkit is provided, which supports three programming modes: incremental mode, composition mode, and parameterization mode. Several types of service sharing are supported by the toolkit. For this, applications are generally assumed to be composed of two parts: inference rules and actions. The incremental and composition mode allow using previously available inference rules for developing new context-aware applications. In the parameterization mode users utilize previously available applications by specifying new parameter values. Each developer introduces its developed inference rule or application as well as required parameters and publishes it to the resource-sharing module. Open provides keyword search and browsing for discovery of previously published inference rules.

- *Privacy protection* - Cobra involves policy-management module, which is responsible for considering user's policies when disseminating their contextual information. Awareness protects privacy of users by following and applying their policies in distributing their contextual information. In SM4ALL, each sensor has a wrapper, which implements access control. Privacy protection has been performed by designing "authorization" as an optional attribute to all context types. It enumerates services that are allowed to access a sensitive context type. In Feel@home, each domain context manager server involves a local access control component, which analyzes context requests against user privacy settings.

**Table II Summary of Functional requirements**

| | | Context Toolkit | Gaia | Cobra | SOCAM | Awareness | SM4ALL | Feel@home Open | CAMEO |
|---|---|---|---|---|---|---|---|---|---|
| **Acquisition** | Discovery | √ | √ | √ | √ | √ | √ | √ | √ |
| | Storage | √ | × | √ | √ | √ | √ | √ | √ |
| **Aggregation in WSN** | | × | × | × | × | × | √ | × | × |
| **QOC** | QOC assessment | × | × | × | × | √ | √ | × | √ |
| | Conflict resolution | × | × | √ | √ | √ | √ | × | × |
| **Modeling** | | √ | √ | √ | √ | √ | √ | √ | √ |
| **Reasoning** | | × | √ | √ | √ | √ | √ | √ | √ |
| **Dissemination** | | √ | √ | × | √ | √ | √ | √ | √ |
| **Service management** | | × | × | × | × | × | × | √ | × |
| **Privacy protection** | | × | × | √ | × | √ | √ | √ | × |

## 3-4 Non-functional attributes

In the following, middleware systems are analyzed for how well each satisfies the quality attributes. At the end, the result of surveying is summarized via table II:

- *Expandability* - None of the surveyed projects supports introducing a new domain to the environment; however, most of them support entity and context introduction. In Context Toolkit, discoverer provides a service for registering context widgets, interpreters and aggregators. This service supports expanding the environment by inserting new context types and entities. Similarly, SOCAM offers context provider and interpreter registration mechanisms. In the Awareness project, it is possible to introduce new entities and context types by registering new context producers. SM4ALL provides context type registration service for inserting new context types. It also provides publisher registry for introducing new context providers. CAMEO provides application registration service, which involves introduction of the application context types.

- *Transparency* - Transparency is considered from two aspects: access and location. In Context Toolkit, applications retrieve their contextual needs in two stages: At first, they issue a query to the discoverer to acquire handle to the widget and then subscribe for the context. This scheme is transparent from both of the aspects. SOCAM provides a uniform query mechanism for context retrieval. After receiving a query, the service locating server performs a semantic matchmaking and returns the reference to the corresponding context provider or interpreter. This scheme is transparent from both of the standpoints. In Awareness, different local context-aware middleware systems utilize different kinds of operators and syntaxes for context dissemination; however, bridges convert these mechanisms to each other. Therefore, users are unaware of this heterogeneity and make use of a uniform operator for context retrieval. Moreover, bridges roughly hide the difference of locations that the contextual information has come from. SM4ALL supports a general structure for queries and users acquire contextual needs by associating a listener to a query for a single context element. The approach is transparent from both of the aspects. Feel@home uses the same query mechanism for retrieving all global contextual information and does not need users to specify domain of the context; therefore, global query mechanism for context dissemination is transparent from both of the aspects. CAMEO provides uniform APIs to users, which are transparent from both of the aspects.

- *Reusability* - Reusability is investigated from two viewpoints. (1) Reusability of context provider (sensing, reasoning, mining, etc) components, and (2) reusability of context-aware service components. Among the surveyed projects only Open framework supports reusability of service components by supporting reusability of inference rule components among application developers. Context Toolkit supports reusability of context provider components through introducing widget concept. SOCAM also supports context provider and interpreter reusability. By introducing the notion of context producer, Awareness supports reusability of context producer components. SM4ALL provides context processors (for aggregation, reasoning, etc) reusability by supporting registration of them to the context processor registry. It also supports context provider components reusability by registering them to the publisher registry. CAMEO also supports reusability of context provider by introducing Device Context Provider, which is in charge of collecting context data derived from internal components of the mobile phone.

- *Reliability* - Among the investigated projects, only Gaia provides a replication scheme for ensuring reliability. An event manager component is designed for decoupling context providers and consumers. If a provider crashes, a replica continues its task to prevent from system crash**.**

**Table III Summary of quality attributes**

|  |  | Context Toolkit | Gaia | Cobra | SOCAM | Awareness | SM4ALL | Feel@home Open | CAMEO |
|---|---|---|---|---|---|---|---|---|---|
| **Expandability** | Domain | × | × | × | × | × | × | × | × |
|  | Entity- Context | √ | × | × | √ | √ | √ | - | √ |
| **Transparency** | Access | √ | - | - | √ | √ | √ | √ | √ |
|  | Location | √ | - | - | √ | √ | √ | √ | √ |
| **Reusability** | Context providing components | √ | × | × | √ | √ | √ | - | √ |
|  | Service components | × | × | × | × | × | × | √ | × |
| **Reliability** |  | × | √ | × | × | × | × | × | × |

## 4. Conclusion

In this chapter, context-aware middleware is investigated by proposing its functional and non-functional requirements and reviewing some well-known projects. According to this study, besides the traditional tasks of middleware, context discovery and storage, aggregation, modeling, reasoning, dissemination, QoC assessment, and service management as well as protecting privacy of the users are the specific tasks that should be supported by context-aware middleware. Moreover, context-aware middleware should satisfy expandability of the environment, transparency from the viewpoint of application programmers, reusability of components, and reliability for users. The most challenging step toward developing context-aware middleware is the architecture design. To support development of various kinds of context-aware applications, a multiple-domain environment should be considered. Designing context-aware middleware for such an environment envisages serious challenges including resource limitations of devices, dynamic nature of the environment, and mobility of entities. In addition to architectural design, there are independent open research directions in most of the other functionalities of the middleware.